\documentclass{article}
\usepackage[utf8]{inputenc}
\usepackage{amsfonts, amsmath, amsthm, amssymb} 
\usepackage{graphicx} 
\usepackage{hyperref} 
\usepackage{multirow,tabu} 
\usepackage{array}
\newcolumntype{H}{>{\setbox0=\hbox\bgroup}c<{\egroup}@{}}
\usepackage{chngcntr}
\usepackage{bbm}
\usepackage{authblk}
\usepackage{color}
\usepackage{float} 
\usepackage[backend=biber,style=authoryear, sorting=nyt,sortcites=true,giveninits=true,maxbibnames=9,natbib ,uniquename=false,maxcitenames=2,uniquelist=false]{biblatex}
\renewbibmacro*{volume+number+eid}{%
  \printfield{volume}%
  \setunit*{\addnbspace}
  \printfield{number}%
  \setunit{\addcomma\space}%
  \printfield{eid}}
\DeclareFieldFormat[article]{number}{\mkbibparens{#1}}
\title{Train performance analysis using heterogeneous statistical models}
\author{Jianfeng Wang\thanks{Corresponding author : Jianfeng Wang\\
\hspace*{4.3 mm} Email: jianfeng.wang@umu.se}}
\author{Jun Yu}
\affil{Department of Mathematics and Mathematical Statistics, Umeå University, SE 901 87 Umeå, Sweden}
\date{}

\begin{document}

\maketitle

\begin{abstract}
This study investigated the effect of harsh winter climate on the performance of high speed passenger trains in northern Sweden. Novel approaches based on heterogeneous statistical models were introduced to analyse the train performance in order to take the time-varying risks of train delays into consideration. Specifically, stratified Cox model and heterogeneous Markov chain model were used for modelling primary delays and arrival delays, respectively. Our results showed that the weather variables including temperature, humidity, snow depth, and ice/snow precipitation have significant impact on the train performance.
\end{abstract}


\section{Introduction}

Coldness, heavy snow and ice/snow precipitation are well known winter phenomenon in the northern region of Sweden. Such climate can cause severe problems to railway transportation as well as people who rely on them, which leads to ineluctable impacts on the normal operations of the whole society. It becomes an especially prominent problem nowadays, as the railway network develops more complicated, the trains run faster, and more people choose railway as their travel mode. The aim of this study is thus to analyse the harsh winter effects on railway operation in northern Sweden. Regarding railway operation, punctuality is one key criterion in order to minimise the society costs and increase reliability of the railway operation. Therefore, the task of the study is to investigate and figure out how train delays are affected by the winter climate.

Primary delay and arrival delay are two commonly used measurements in the train operation. Primary delay measures the increment in delay within two consecutive measuring spots in terms of running time, and arrival delay is the delay in terms of arrival time at a measuring spot. The time limits to define primary delays and arrival delays vary from country to country \citep{Yuan2007}. According to Swedish Transport Administration (STA), a train arriving at one measuring spot within five minutes is not considered as arrival delay and a delay of three minutes or more in terms of running time within two
consecutive measuring spots is considered as primary delay. One of the main interests from STA is to investigate how the two kinds of train delays are affected by winter weather. Therefore, we apply the STA criteria throughout the study.

A number of studies about train performance analysis have been conducted. \citet{Yuan2007} used probability models based on blocking time theory to estimate the knock-on delays of trains caused by route conflicts and late transfer connections in stations. In \citet{MURALI2010483}, the authors modeled travel time delay as a function of the train mix and the network topology. \citet{LESSAN20191214} proposed a hybrid Bayesian network model to predict arrival and departure delays in China. \citet{Huang8883805} pointed out in their paper that arrival delay was highly correlated to capacity utilization of the train line. In a more recent study, \citet{HUANG2020338} applied Bayesian network to predict disruptions and disturbances during train operations in China. In addition to those, a few earlier studies about relations between train performance and weather effects have also been investigated. \citet{Xia2013} fitted a linear model and showed that weather variables like snow, temperature, precipitation and wind had significant effects on the punctuality of trains in Netherlands. \citet{BRAZIL201769} used simple multiple linear regression model and demonstrated that weather variables, such as wind speed and rainfall, can have a significantly negative impact on arrival delays in Dublin area rapid transit rail system. A machine learning approach was used to create a predictive model to predict the arrival delay at each station for a train line in China with help of weather observations in \citet{Wang2019}. \citet{Ottosson} used negative binomial regression and zero-inflated model and showed that weather variables, such as snow depth, temperature and wind direction, had significant effects on the train performance. A recent study by \citet{wang2020effects} applied non-stratified Cox model and homogeneous Markov chain model to analyse the weather effects on the primary delay and arrival delay, respectively. The authors treated primary delay as recurrent time-to-event data, and the transitions between states (arrival) delay and punctuality in a train trip as a Markov chain. One limitation is that the hazard function in the Cox model was assumed to be constant over events and the transition intensity in the Markov chain model can not change at any specified time. However, these assumptions are often not realistic.
 
 In this study, we relax the restrictions in \citet{wang2020effects} by assuming the heterogeneity in the models, i.e. hazard functions vary among events and transition intensity may change at any specified time point. The main contribution is that we prove that the heterogeneous models outperform the homogeneous counterparts. In addition, to the best of our knowledge, it is the first study to apply the heterogeneous models to investigate the weather impacts on the train delay issues, i.e. a stratified Cox model is used to investigate how the winter climate affects the occurrence of primary delays, and a heterogeneous Markov chain model is applied to study the effect of winter climate on the transitions between delayed and punctual states. 

The paper is organized as follows. In Section 2, we introduce the statistical models in details. Data processing and analysis methods are described in Section 3. Section 4 is reserved for results. Section 5 is devoted to the conclusion and discussion.

\section{Statistical modelling}
In the section, the two statistical models, i.e. stratified Cox model and heterogeneous Markov chain model, are introduced in details. 
\subsection{Stratified Cox model with time dependent covariates for recurrent event}
As an extension of original Cox
models in \citet{Cox,Andersen1982}, \citet{PRENTICE} proposed a stratified Cox model, which is  commonly used for modelling recurrent events in survival analysis. It will be used in this study to analyse the relationship between hazards of trains with recurrent event (primary delay) and weather covariates by assuming that the hazard function of a train is correlated to its preceding events through an event-specific baseline hazard function. Formally, the stratified Cox model with time dependent covariates for recurrent event is an expression of the hazard function and covariates

\begin{equation}
\label{coxmodel}
   h_{ij}(t)  = h_{0j}(t)\exp{(\boldsymbol{\beta}^T \mathbf{x}_{ij}(t))},
\end{equation}
where 
\begin{itemize}
\item $h_{ij}(t)$ represents the hazard function for the $j$th event of
the $i$th train at time $t$.
\item $h_{0j}(t)$ is an event-specific baseline hazard and the order number $j$ is the stratification variable, e.g. $h_{01}(t)$ is a common baseline hazard of the first event for each train.
\item $\mathbf{x}_{ij}(t)$ represents weather covariate vector for the $i$th train and the $j$th event at time $t$.
\item $\boldsymbol{\beta}$ is an unknown  coefficient vector to be estimated, exponential of which indicates how the hazard ratios are affected by the covariate vector.
\end{itemize}

The coefficients can be estimated by maximising the partial likelihood, given by  
\begin{equation}
    \label{partialLikli}
    L(\boldsymbol{\beta})  = \prod_{i = 1}^n \prod_{j = 1}^{k_i} \left(\frac{\exp{( \boldsymbol{\beta}^T \mathbf{x}_{i}(t_{ij})))}}{\sum_{l \in R(t_{ij})} \exp{( \boldsymbol{\beta}^T \mathbf{x}_{l}(t_{ij})})}\right)^{\delta_{ij}},
\end{equation}
where $j$ is the event index with $k_i$ being the train-specific maximum number of events, $\mathbf{x}_{i}(t_{ij})$ denotes the covariate vector for the $i$th train at the $j$th event time $t_{ij},$ $\delta_{ij}$ is an event indicator which equals $1$ for the $j$th event of the $i$th train and $0$ for censoring, $R(t_{ij}) = \{l, l=1,\cdots,n: t_{l(j-1)}< t_{ij}\leq t_{lj}\}$ is a group of trains that are at risk for the $j$th event at time $t_{ij}$. Note that the partial likelihood takes into account the conditional probabilities for the events that occur for trains.

The fitted model can then be used to predict the hazard function, $\hat{h}_{ij}(t)$, for the $j$th event of train $i$ of interest given the values of covariates, as well as corresponding survival function, $\hat{S}_{ij}(t)$, which gives the probability that train $i$ does not suffer the $j$th event up to time $t$. The survival function is exponential function of the hazards function, i.e. $\hat{S}_{ij}(t)=\exp\left(-\int_0^t \hat{h}_{ij}(x)\, \mathrm{d}x\right)$. 

\subsection{Heterogeneous Markov chain model with time dependent covariates}
Let $\{Y(t), t\ge 0\}$ denote a continuous time Markov chain. At each time point $t$, $Y(t)$ takes a value over a countable state space. The probability of chain $Y(t)$ being in state $s$ at time $t$ is $P(Y(t)=s)$. The conditional probability $p_{rs}(t, t+u)=P(Y(t+u) = s | Y(t) = r) $ represents the transition probability of moving from the state $r$ at time $t$ to the state $s$ at time $t+u$. The instantaneous movement from state $r$ to state $s$ at time $t$ is governed by transition intensity, $q_{rs}(t)$, through the transition probabilities
\begin{equation}
\label{inte}
    q_{rs}(t) = \lim_{\Delta t \to 0} P( Y(t+\Delta t) = s |Y(t)=r)/\Delta t.
\end{equation}
With these definitions, a Markov chain can be used to describe train running states (delay/punctuality) on a train line, where the time $t$ refers to running distance of a train from the starting point throughout the study instead of time, since the running distance is more meaningful in practice. The $q_{rs}(t)$ of a $q$ states process forms a $q\times q$ transition intensity matrix $Q(t)$, whose rows sum to zero, so that the diagonal entries are defined by $q_{rr}(t) = -\sum_{s\neq r}q_{rs}(t)$. An example of transition intensity matrix $Q(t)$ with two states can be seen below

\begin{equation}
Q(t) = \begin{bmatrix}
    q_{11}(t) & q_{12}(t) \ \\
    q_{21}(t) \ & q_{22}(t) \end{bmatrix},
\end{equation}
where $q_{11}(t)=-q_{12}(t)$ and $q_{22}(t)=-q_{21}(t)$ at time $t$.

A homogeneous Markov chain in time means that the transition intensity $Q(t)$ is independent of $t$, and the transition probability from one state to another depends solely on the time difference between two time points, i.e. 

\begin{equation}
    \label{mark2}
    P( Y(t+u) = s |Y(t) = r) = P(Y(u) = s | Y(0) = r).
\end{equation}

Corresponding to the transition intensity matrix $Q$, the entry in a transition probability matrix $P(t,t+u)$ is the transition probability $p_{rs}(t,t+u)$. The relationship between transition intensity matrix and transition probability matrix is specified through the Kolmogorov differential equations \citep{Cox1977}. Specially, when a process is homogeneous, the transition probability matrix can be calculated by taking the matrix exponential of the transition intensity
matrix 
\begin{equation}
\label{pro}
    P(t,t+u)=P(u) = \text{Exp}(uQ).
\end{equation}

In a homogeneous Markov chain model, to take account of the effect of covariates, a Cox like model was proposed by \citet{Marshall1995}
\begin{equation}
\label{explo}
    q_{rs} = q_{rs}^{(0)} \exp{(\boldsymbol{\beta}_{rs}^T \mathbf{x}_{rs})},
\end{equation}
where $q_{rs}^{(0)}$ is a baseline transition intensity from state $r$ to state $s$ when all covariates are zero and $\mathbf{x}_{rs}$ is a covariate vector under the corresponding transition.  The value $\exp{(\beta_{rs})}$, where $\beta_{rs}$ is one element of the vector $\boldsymbol{\beta}_{rs},$ reflects how the corresponding covariate affects the hazard ratio given that all other covariates are held constant. More specifically, $\exp{(\beta_{rs})}>1$ indicates the transition intensity from $r$ to $s$ increases as the value of the covariate increases, $\exp{(\beta_{rs})}<1$ indicates the transition intensity decreases as the value of the covariate increases, while $\exp{(\beta_{rs})}=1$ implies the covariate has no effect on the transition intensity.

The coefficient vectors $\boldsymbol{\beta}_{rs}$ as well as the transition intensity matrix $Q$ and the transition probability matrix $P(t)$ can be estimated through maximising the likelihood 

\begin{equation}
    \label{logts}
    L(Q) =  \prod_{i = 1}^n \prod_{j = 1}^{c_i} p_{Y(t_{i,j}),Y(t_{i,j+1})}(t_{i,j + 1}- t_{i,j}),
\end{equation}
where $j$ is a sequence index of observed states with $c_i$ being number of measuring spots for train $i$ on the train line, $Y(t_{i,j})$ represents the $j$th observed state of the $i$th train at time $t_{i,j}$ and the transition probability is evaluated at the time difference $t_{i,j + 1}- t_{i,j}$. 

Contrary to homogeneous Markov chain model, a heterogeneous Markov chain model assumes that the transition intensity may change continuously at any time. However, the transition probability matrix as well as the likelihood (\ref{logts}) are analytically intractable under this situation \citep{Titman}. An exception is that the transition intensity changes at countable time points. For example, the transition intensity is assumed to change at time point $t_0$ for each train. To achieve it, one can introduce an indicator covariate in the model to represent the two time periods
\begin{equation}
\label{exten}
    q_{rs}(t) = q_{rs}^{(0)} \exp{(\boldsymbol{\beta}_{rs}^T \mathbf{x}_{rs}^{(\mathbbm{1}_{\{t\ge t_0\}})}+z_{rs}\mathbbm{1}_{\{t\ge t_0\}})},
\end{equation}
where $\mathbbm{1}$ is an indicator function taking value 1 if $t\ge t_0$, otherwise, 0, and $z_{rs}$ is the coefficient. Note that the covariate vector under the same transition is separated into two at $t_0$ through the indicator function, since (\ref{exten}) can be formulated as two homogeneous models and each model has its own covariate vector, i.e. $\mathbf{x}_{rs}^{(0)}$ for the first model when $t<t_0$ and $\mathbf{x}_{rs}^{(1)}$ for the second model when $t\ge t_0$.  Similar to $\exp{(\beta_{rs})}$, the value $\exp{(z_{rs})}$ is the hazard ratio of intensities between $t\ge t_0$ and $t<t_0$ for the transition from $r$ to $s$. 

After fitting the heterogeneous Markov chain model, one can calculate the predicted transition probability matrix for any operational interval of interest on the train line using (\ref{pro}) provided that values of covariates for the interval are given.

\section{Method}
This section describes the train data and weather data used for the analysis as well as the missing data imputation method for train data. Besides, a models comparison method,  likelihood ratio test,  is presented which is used to compare the performance between the heterogeneous models and homogeneous models.

\subsection{Train data}
Our investigation focuses on high speed passenger trains, which is a type of trains with top speed of between 200 to 250 km/h, between Umeå and Stockholm in the northern region. The high speed passenger train is chosen, because this type of train has higher priority on the train line and often travels longer distances, which can minimises non-natural effects on the train line so that it is easier to detect the pure weather impacts. The data window chosen is December 2016 - February 2017, which is typical winter time in Sweden.
  
A train line comprises of a number of measuring spots where the operational times are recorded such as departure and arrival times. The train line between Umeå and Stockholm includes 116 measuring spots in total. The total length of the train line is 711 km and the planned drive time for a high speed passenger train is approximately 6.5 hours. The lengths of any two consecutive measuring spots vary from 0.3 km to 15 km. The key variables are listed in {Table \ref{table:1}}.
\begin{table}[H]
\centering
\caption{List of variables in the train operation data}
\medskip
\begin{tabular}{  m{11em} | m{11cm}  } \hline\hline
\textbf{Variables}  & \textbf{Description} \\ \hline
Train Number & An identification number for train used in the trip  \\ \hline
Arrival location & Name of arrival measuring spot  \\ \hline
Departure location & Name of departure measuring spot \\ \hline
Departure date &  The departure date (yyyy-mm-dd) for a train at a location.  \\ \hline
Arrival date &  The arrival date (yyyy-mm-dd) for a train at a location. \\ \hline
Train type &  Type of train, for example: high speed, commute train and regional \\ \hline
Section Length  & Length (km) between two consecutive measuring spots \\ \hline
Planned departure time  & The planned departure time (hh:mm) at a measuring spot \\ \hline
Planned arrival time  & The planned arrival time (hh:mm) at a measuring spot \\ \hline
Actual departure time  & The Actual departure time (hh:mm) at a measuring spot\\ \hline
Actual arrival time  & The Actual arrival time (hh:mm) at a measuring spot \\ \hline\hline
\end{tabular}
\label{table:1}
\end{table}

\begin{figure}[H]
    \centering
    \includegraphics[scale=0.4]{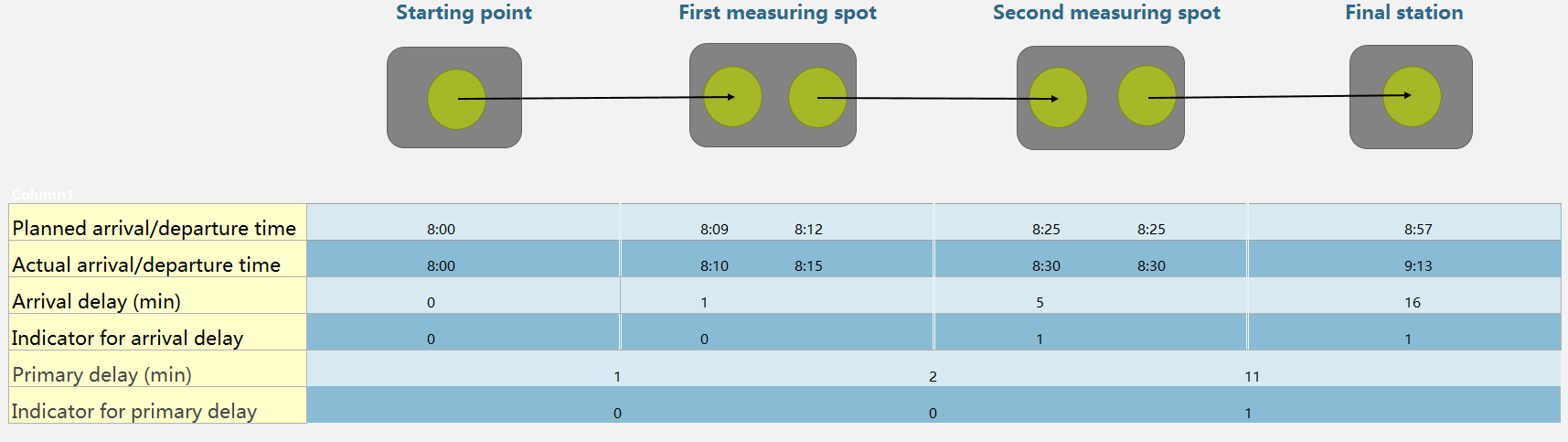}
    \caption{Illustration of a train run with derived indicators for primary and arrival delays}
    \label{pic:timetable}
\end{figure}

To fit the two statistical models to the train data, the data should be organized to include the following variables, e.g. each record has one departure spot of the train run (it is not necessary in the Markov chain model), its subsequent arrival spot, distances of these two measuring spots from starting station, and indicators of primary delay and arrival delay, 0/1, for this running section, as well as corresponding weather covariates and train identification number. To obtain the indicator variables for primary delay and arrival delay, one needs calculate the running time difference and arrival time difference compared to the schedule, which are (Actual arrival time$-$Actual departure time)$-$(Planned arrival time$-$Planned departure time)  and (Actual arrival time$-$Planned arrival time), respectively. Afterwards, the values for the two indicator variables can be assigned, i.e. 1 stands for a primary/arrival delay, 0 otherwise. An example about how to derive the indicator variables along a train line is illustrated in {Figure \ref{pic:timetable}}.

\subsection{Weather data} 
\begin{figure}[H]
    \centering
    \includegraphics[scale=0.7]{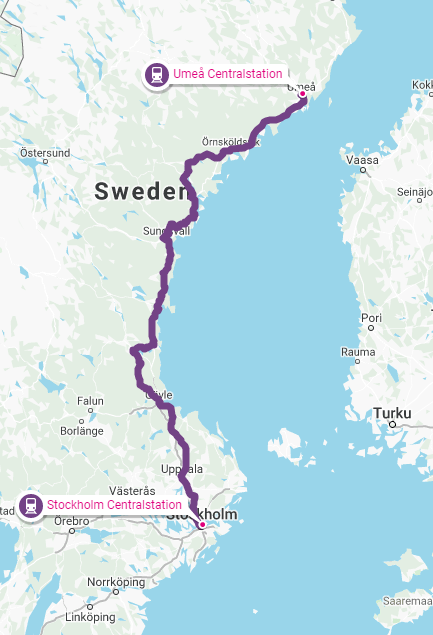}
    \caption{Train line in the region with simulated WRF data}
    \label{pic:weather}
\end{figure}

 The weather data from December 2016 to February 2017 is simulated from the Weather Research and Forecasting (WRF) model instead of using real meteorological observations, since the distances between the nearest meteorological station and measuring spot along the train line range from 17 to 24 km \citep{Ottosson}. Thus, using the meteorological data is not an ideal choice in the analysis. However, a WRF model is a numerical weather prediction system that is used for research and operational purposes. Its reliable performance has been assessed in a number of studies \citep{wangbayesian, wangdownscal,Mohan,Cassano}. The WRF model simulates desired weather variables estimations over grids. Higher spatial resolution implies smaller grids over a region of interest. Temporal resolution decides the time interval between each simulation. Therefore, a WRF with high spatio-temporal resolution is a good alternative under this situation. In this study, the spatial resolution is set as $3\times3$ km and the temporal resolution is set as 1 hour. The simulation region as well as the train line of interest are shown in {Figure \ref{pic:weather}}.

The weather variables of interest are shown in {Table \ref{table:2}}. These variables are chosen because they are believed to have impacts on the train operation in winter and have been used in \citet{wang2020effects,Ottosson}.

\begin{table}[H]
\centering
\caption{The weather variables of interest}
\medskip
\begin{tabular}{ m{10em} | m{10cm}  } 
\hline\hline
\textbf{Variables}  & \textbf{Description} \\ 
\hline
Temperature  & The temperature ($^\circ$C) at 2 meters above the ground \\  \hline 
Humidity  & Relative Humidity (\%) at 2-meters   \\  \hline
Snow depth & The snow depth in centimeters (cm)  \\  \hline
Ice/snow precipitation  &  Hourly accumulated ice/snow  in millimeter (mm) \\  \hline\hline

\hline
\end{tabular}

\label{table:2}
\end{table}
 The measuring time in train operation data has to be rounded to the closest hour, so that every measuring spot on the train line can be matched with the closest grid point by date and time. 
 
The average of the weather variables within any two consecutive spots are calculated and used in the analysis. Since a large number of the ice/snow precipitation values are zero along the train line, a categorical variable is used instead of the continuous variable, i.e. 0 if ice/snow precipitation is zero, 1 otherwise.
 
\subsection{Missing values in the train operation data}
A section between two consecutive measuring spots for a train trip often has missing departure/arrival times that can be classified into three different classes which are defined in {Table \ref{table:miss}}.

\begin{table}[H]
\centering
\caption{Classes of missing times}
\medskip
\begin{tabular}{  m{8em} | m{5cm}| m{5cm}   } 
\hline\hline
\textbf{Class}  & \textbf{Departure time missing}& \textbf{Arrival time missing} \\ 
\hline
1& True & False\\  \hline 
2& False & True \\  \hline 
3& True & True \\  \hline \hline
\end{tabular}

\label{table:miss}
\end{table}

 A common method to impute missing values in such longitudinal data is called last observation carried forward (LOCF), i.e. the latest recorded value is used to impute the missing value.  The advantages of using LOCF are that the number of observations removed from the study decreases and make it possible to study all subjects over the whole time period. A disadvantage with the method is the introduction of bias of the estimates if the values changes considerably large with time or the time period between the most recent value and the missing value is long. 
Because the intervals with missing values are short in the dataset which decreases the risk of bias, thus it is reasonable to apply this approach. Based on the LOCF, the imputation procedure is explained further below.

\begin{enumerate}
  \item Start from the beginning of the trip and save the latest arrival and departure time until a missing time is occurring. 
  
  \item If the missing time is arrival time then (a); if departure time is missing then (b):
  \begin{enumerate}
     \item Replace the missing arrival time with the latest departure time + the planned driving time for the previous section
     \item Replace the missing departure time with the latest arrival time  + the planned dwell time. 
     \end{enumerate}
     
    \item Save the imputed time as the new latest time.
    
    \item If the section is not the last section of the trip, go back to step 1.
    
\end{enumerate}

\subsection{Likelihood ratio test}
The likelihood ratio test is a hypothesis test that helps to determine whether adding complexity to a simple model makes the complex model significantly better compared to the simple model. Under the study context, comparisons occur between the two (complex) heterogeneous models against the two (simple) homogeneous models, respectively. The likelihood ratio test statistic is given by
\begin{equation}
\label{lr}
    \lambda = -2\ln \left(\frac{\mathcal{L}_{homo}(\hat{\boldsymbol{\theta}})}{\mathcal{L}_{heter}(\hat{\boldsymbol{\theta}})}\right),
\end{equation}
where the numerator in the bracket is the likelihood value for a homogeneous model with estimated parameter vector $\hat{\boldsymbol{\theta}}$, while the denominator represents the likelihood for the corresponding heterogeneous model. The null hypothesis is the simple model is better and a low $p$ value leads to the rejection of the null hypothesis and in flavour of the complex model. 

\subsection{Analysis tool}
R is the software used for data processing and modelling. Specifically, the package \textit{survival} is used for the stratified Cox model and the package \textit{msm} is used for the heterogeneous Markov chain model. 

\section{Results}
\subsection{Stratified Cox model}
The estimates from the fitted stratified Cox model with 95\% confidence intervals (CIs) and $p$-values can be found in \textit{Table \ref{table:solocox}}. Temperature and humidity are the two variables that have significant effects on the occurrence of the primary delay. To be specific, as temperature increases with $1^\circ$C, the hazard decrease 3.6\%, and  as humidity increase 1\%, the hazard increases 1.7\%. Comparison between the stratified Cox model and the non-stratified Cox model in \citet{wang2020effects} by using a likelihood ratio test shows that the stratified model is significantly better than the non-stratified model ($p<0.0001$).

\begin{table}[H]
\caption{Estimates from the fitted stratified Cox model }
\medskip
\centering
\begin{tabular}{  m{10em} |  m{2.5cm}| m{2cm}| m{2cm}|    m{2cm} } 
\hline\hline
\textbf{Predictor}  &\textbf{Hazard ratio} & \textbf{CI: Lower}& \textbf{CI: Upper} &  \textbf{$p$-value}\\ \hline
Temperature  & 0.964&0.933&0.997& 0.0338     \\\hline 
Humidity & 1.017&1.003&1.030&  0.0107    \\\hline 
Snow depth & 1.017&0.990&1.044 &0.2197     \\\hline 
Ice/snow precipitation & 1.055&0.841&1.323 &0.6465    \\\hline \hline

\end{tabular}

\label{table:solocox}
\end{table}

\begin{figure}[H]
    \centering
    \includegraphics[scale=0.6]{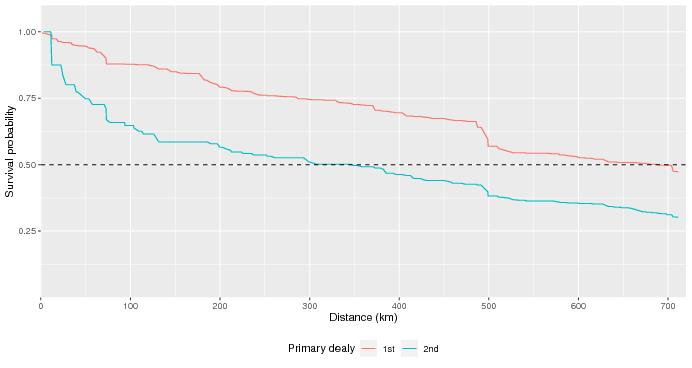}
    \caption{Survival probabilities for the first two primary delays}
    \label{pic:sur}
\end{figure}

Besides hazard ratios, a survival plot is also produced to show how survival probabilities vary between the first and second occurrence of primary delays in {Figure \ref{pic:sur}}. The survival curves for the higher orders of primary delays are not shown due to the data deficiency. The curves are plotted under the condition with the average of temperature, humidity and snow depth among the whole data together with ice/snow precipitation, i.e. temperature is $-1.2^\circ$C, humidity is 85\%, snow depth is 3 cm and ice/snow precipitation is 1.

The figure indicates clearly that slightly less than 50\% of trains do not experience any primary delay during the trip, and 50\% of the trains which have experienced the first primary delay suffer the second primary delays after running 330 km from the starting point. It is interesting to notice that there is a substantial reduction in survival probability right before running 500 km from the starting point for the first primary delay. The reason might be related to some mechanical problems of trains in winter. 

\subsection{Heterogeneous Markov chain model}
As indicated in {Figure \ref{pic:sur}}, under the average weather condition, half of the number of the first two primary delays occur in the second part of the trip partitioned at 330 km, thus it is reasonable to assume the transition intensities are different before and after running 330 km. Therefore, $t_0=330$ is chosen in modelling heterogeneous Markov chain (\ref{exten}). {Table \ref{table:markov}} and {\ref{table:markov2}} present the hazard ratios from the heterogeneous Markov chain model with 95\% CIs and $p$-values. The ice/snow   precipitation has significant impact on the transition from punctual to delayed states in {Table \ref{table:markov}}, which means that the transition intensity from punctuality to delay increases 23\% with ice/snow precipitation. In contrast, temperature, humidity and snow depth have significant impacts on the transition from delayed to punctual states. It indicates in {Table \ref{table:markov2}} that as the temperature increases 1 $^\circ$C, the transition intensity from delayed to punctual states increases 4.4\%, as the humidity increases 1\%, the transition intensity decreases 1.6\%, and as the snow depth increases 1 cm, the transition intensity decreases 4.8\%. Likelihood ratio test between the heterogeneous Markov chain model  (\ref{exten}) and the homogeneous Markov chain model in \citet{wang2020effects} is also performed, which shows that our new model  (\ref{exten}) fits significantly better with $p<0.0001$ than the homogeneous one.

\begin{table}[H]
\centering
\caption{Hazard ratios from punctual to delayed states}
\medskip
\begin{tabular}{  m{10em} | m{2.5cm}|  m{2cm}| m{2cm}|m{2cm}} 
\hline\hline
\textbf{Predictor}  & \textbf{Hazard Ratio} & \textbf{CI: Lower}& \textbf{CI: Upper}& \textbf{$p$-value}   \\ \hline
Temperature & 0.988&0.962&1.015&0.3881  \\\hline 
Humidity &  0.990 &0.980&1.001&0.0683 \\\hline 
Snow depth & 1,000& 0.977&1.026 &0.9442 \\\hline 
Ice/snow precipitation & 1.230&1.001&1.512 &0.0489  \\\hline \hline

\end{tabular}

\label{table:markov}
\end{table}

\begin{table}[H]
\centering
\caption{Hazard ratios from delayed to punctual states}
\medskip
\begin{tabular}{  m{10em} | m{2.5cm}|  m{2cm}| m{2cm}|m{2cm}} 
\hline\hline
\textbf{Predictor}  & \textbf{Hazard Ratio} & \textbf{CI: Lower}& \textbf{CI: Upper}&\textbf{$p$-value}    \\ \hline
Temperature & 1.044&1.016&1.073 &0.0022 \\\hline 
Humidity &  0.984 &0.973&0.995&0.0036 \\\hline 
Snow depth & 0.952& 0.924&0.980 &0.0012 \\\hline 
Ice/snow precipitation  & 0.890&0.719&1.103&0.2921   \\\hline \hline

\end{tabular}

\label{table:markov2}
\end{table}

\begin{figure}[H]
    \centering
    \includegraphics[scale=0.6]{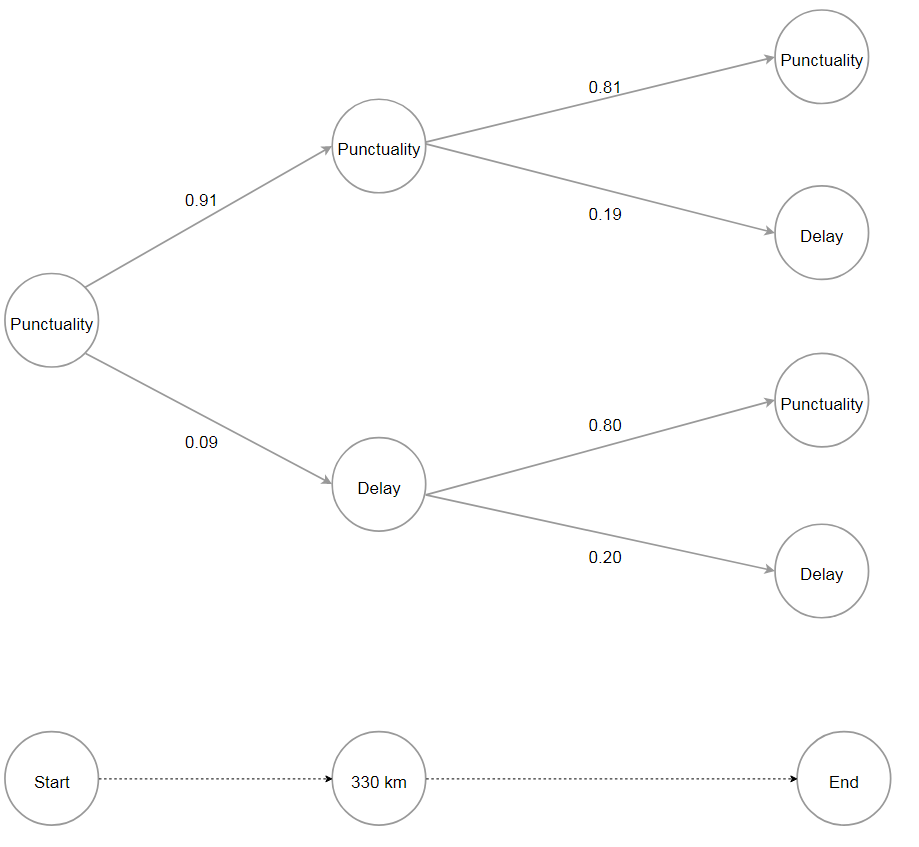}
    \caption{Probabilities of evolution of delayed status in the trip}
    \label{pic:prob}
\end{figure}

By using the average of temperature, humidity and snow depth together with ice/snow precipitation, {Table \ref{table:intens}} and {Figure \ref{pic:prob}} show the estimated transition intensities and probabilities of evolution of delayed status for the two segments divided at 330 km, respectively. In {Table \ref{table:intens}}, the transition intensity of the second segment of the trip from punctual to delayed states is 58.6\% higher than the first, however, transition intensity of the second segment of the trip from  delayed to punctual states is 31.9\% lower than the first segment. In other words, the second segment is much easier to suffer delay and more difficult to recover from a delay. It is also verified in {Figure \ref{pic:prob}} that the first segment has higher probability to be punctual. Overall, the probability of a train arriving the final station on time is about 80\%.
\begin{table}[H]
\centering
\caption{Estimated Hazard ratios between segments $[330, \text{end})$ and $[0,330)$ }
\medskip
\begin{tabular}{m{9em} | m{2.5cm}|  m{2cm}| m{2cm}| m{2cm}} 
\hline\hline
\textbf{Predictor}  & \textbf{Hazard Ratio} & \textbf{CI: Lower}& \textbf{CI: Upper}& \textbf{$p$-value}    \\ \hline
Punctuality - delay & 1.586&1.337&1.883& $<0.0001  $ \\\hline 
Delay - punctuality & 0.681&0.564&0.822&$<0.0001$   \\\hline \hline
\end{tabular}

\label{table:intens}
\end{table}

\section{Conclusion and discussion}

This study investigated the effect of harsh winter climate on the performance of high speed passenger trains in northern Sweden, with respect to the occurrence of primary delays and the transition intensities between delayed and punctual states. Novel approaches based on heterogeneous statistical models were introduced to analyse the train performance in order to take the time-varying risks of train delays into consideration. Specifically, stratified Cox model and heterogeneous Markov chain model were used for modelling primary delays and arrival delays, respectively. We conclude that 1) the two heterogeneous models outperform the homogeneous counterparts; 2) the weather variables, including temperature, humidity, snow depth, and ice/snow precipitation, have significant impact on the train delays. 

In the study, we considered the heterogeneity within each train in both two statistical models, however, the heterogeneity among trains are not touched yet and could be considered in the further investigation, for example frailty Cox model and/or fitting the two models from Bayesian perspective with random effects among trains \citep{vanniekerk2019new}. In addition, how to choose the changing point and how many changing points in a  heterogeneous Markov chain process become critical problems, since the estimated transition intensity matrix may be sensitive to the choices which are very subjective. In this study, only one changing point at 330 km was used, which was decided by the fact that half of the number of the first two primary delays occurred in the second part of the trip partitioned at 330 km under the average weather condition in {Figure \ref{pic:sur}}. Continuously changing transition intensities, which are smooth function of time, for example, Weibull distributed time function, may be more plausible with the help of numerical approximation methods \citep{Titman}. Besides, more could be done in terms of statistical modelling. For instance, 1) a more than two states' Markov chain model can be used to acquire a deeper understanding about the climate effects; 2) more than one changing point of the transition intensity can be investigated in the model; and 3) interactions between weather variables and the indicator variable could be considered to account for the weather effects in each segment in the heterogeneous Markov chain model. Besides, train operation data from more than one winter could to be included in the model fitting procedure to acquire more robust inference.

\section*{Acknowledgements}
We acknowledge EU Intereg Botnia-Atlantica Programme and Regional Council of V\"{a}sterbotten and Ostrobothnia for their support of this work through the NoICE project. We would like to thank the Swedish Transport Administration for providing the train operation data, the Atmospheric Science Group at Lule\r{a} University of Technology for providing the WRF data, and the High Performance Computing Center North (HPC2N) and the Swedish National Infrastructure for Computing (SNIC) for providing the computing resources needed to generate the WRF data.  
  
\printbibliography
\end{document}